\documentclass[
  aps,
  prb,
  reprint,
  superscriptaddress,
  longbibliography,
  showkeys,
  citeautoscript,
]{revtex4-2}
\usepackage[intlimits]{amsmath}
\usepackage{amssymb}
\usepackage{amsthm}
\usepackage{booktabs}
\usepackage{braket}
\usepackage{color}
\usepackage{graphicx}
            
\usepackage[colorlinks=true,
            linkcolor=black,
            citecolor=black,
            filecolor=black,
            urlcolor=black
            ]{hyperref}
\usepackage[utf8]{inputenc}
\usepackage[version=4]{mhchem}
\usepackage{multirow}
\usepackage{soul}
\usepackage{tabularx}
\usepackage{units}
\usepackage[dvipsnames]{xcolor}

\usepackage{dcolumn}
\usepackage{booktabs}
\usepackage{miller}

\usepackage[T5,T1]{fontenc}

\newcommand{\dtu}{
    Department of Electrical and Photonics Engineering, Technical University of Denmark,
    2800 Kgs. Lyngby, Denmark
}

\newcommand{\NanoPhoton}{
    NanoPhoton - Center for Nanophotonics, 
    Technical University of Denmark,
    2800 Kgs. Lyngby, Denmark
}

\makeatletter
\newcommand\emailx[1]{%
\move@AF%
\def\@affil{{\normalfont\,#1\strut}{}}%
}%

\begin{document}

\title{Toward triggered generation of indistinguishable single-photons from MoTe$_{2}$ quantum emitters}

\author{Paweł Wyborski}
\affiliation{\dtu}

\author{Athanasios Paralikis}
\affiliation{\dtu}

\author{Pietro Metuh}
\affiliation{\dtu}

\author{Martin A. Jacobsen}
\affiliation{\dtu}

\author{Christian C. Ruiz Madera}
\affiliation{\dtu}
\affiliation{\NanoPhoton}

\author{Niels Gregersen}
\affiliation{\dtu}

\author{Battulga Munkhbat}
\email{bamunk@dtu.dk}
\affiliation{\dtu}

\begin{abstract}

Single-photon sources operating at telecom wavelengths are fundamental components for long-distance optical quantum communication and information processing. Two-dimensional (2D) transition metal dichalcogenides (TMDs) offer a promising platform for such sources, but their development has been hindered by limited spectral range and poor single-photon indistinguishability. Here, we demonstrate a reproducible and systematic approach for generating near-infrared (1090–1200~nm) quantum emitters in bilayer MoTe$_2$ using deterministic strain and defect engineering. These emitters exhibit strong linear polarization (DOLP $>70\%$), sub-nanosecond lifetimes ($\tau \sim$~130–450 ps), high single-photon purity with triggered $g^{(2)}(0)$ values as low as $\sim$0.01 ($\sim$0.16) under p-shell (quasi-resonant) excitation, and resolution-limited emission ($\sim$150~\textmu eV). Electrostatic biasing enables tuning over a $\sim$3 meV range, suppresses photon bunching, and significantly shortens radiative lifetimes, yielding narrow emission with ratios of experimental to transform-limited linewidths as low as $R\sim55$.
Most notably, two-photon interference measurements reveal a Hong-Ou-Mandel visibility of $V_\text{HOM}\sim$~7.1$\%$ (3.6$\%$), and up to $V_\text{HOM}\sim$~60$\%$ ($\sim$40$\%$) with post-selection by temporal filtering under p-shell (quasi-resonant) excitation. To our knowledge, this presents the highest reported indistinguishability for TMD quantum emitters and the first such demonstration for MoTe$_2$ platform. 
These results establish MoTe$_2$ as a viable platform for tunable, low-noise, high-purity single-photon sources with state-of-the-art indistinguishability for TMD quantum emitters, paving the way for their integration into telecom-compatible quantum photonic technologies.

\end{abstract}

\maketitle

\section{Introduction}

The generation of on-demand, indistinguishable single photons is essential for advancing photonic quantum technologies, enabling quantum interference, entanglement distribution, and scalable quantum computing architectures \cite{Vajner2022Quantum, Maring2024AVersatile}. Operating at longer wavelengths, particularly within the telecom bands, is highly advantageous as it leverages existing low-loss optical fiber infrastructure and ensures compatibility with classical communication networks, thereby enabling long-distance quantum communication with minimal loss \cite{Yu2023Telecom, Holewa2025Solid}.\par
Two-dimensional (2D) transition metal dichalcogenides (TMDs) have emerged as a promising platform for quantum photonics \cite{Gao2023Atomically}, offering efficient photon extraction due to their atomically thin geometries \cite{Brotons2018Engineering}, along with tunable emission and strong light–matter interactions \cite{Wang2018Colloquium, Stuhrenberg2018Strong, schroder2025strong, Han2025Infrared}. Their innate ability to form van der Waals heterostructures enables the engineering of novel optical properties and facilitates integration with existing photonic platforms \cite{Peyskens2019Integration, Errando2021Resonance, Montblanch2023Layered}. Among their many advantages, 2D TMDs can host localized quantum emitters that operate as single-photon sources, with emission energies tunable through strain, defect engineering, and electrostatic control \cite{tonndorf2015single, kumar2015strain, palacios2017large, branny2017deterministic, Tripathi2018Spontaneous, Rosenberger2019Quantum, chakraborty2019electrical, Parto2021Defect, Stevens2022Enhancing, Kumar2024Strain, Wu2025Modulation}. Localized quantum emitters in TMDs are typically created by coupling local strain fields with defect sites \cite{Parto2021Defect, Linhart2019Localized, Kumar2024Strain, Wu2025Modulation}, giving rise to confined excitonic states that exhibit quantum light emission. These emitters have been demonstrated across multiple TMD materials and spectral ranges \cite{Montblanch2023Layered}, with most reports focusing on tungsten-based compounds such as WSe$_2$ operating in the visible regime \cite{He2015Single, srivastava2015optically, kumar2015strain, tonndorf2015single, Koperski2015Single, Shepard2017Nanobubble, branny2017deterministic, palacios2017large, Luo2018Deterministic, Parto2021Defect, Wang2021Highly, Sortino2021Bright, Drawer2023Monolayer, Tripathi2018Spontaneous, paralikis2024tailoring, Montblanch2023Layered}. While these systems can achieve high single-photon purity \cite{Kumar2016Resonant, branny2017deterministic, Daveau2020Spectral, Parto2021Defect, Drawer2023Monolayer, vonHelversen2023Temperature, piccinini2024high} and tunability \cite{chakraborty2019electrical, kim2019position,  lenferink2022tunable, Paralikis2025Tunable}, their limited indistinguishability of only $\sim$2$\%$ \cite{Drawer2023Monolayer} and short emission wavelengths constrain their use in long-distance quantum networks. \par
Molybdenum-based TMDs, such as MoTe$_2$, offer emission in the near-infrared (NIR), overlapping with telecom-relevant windows. However, quantum emitters in MoTe$_2$ have remained largely unexplored, with previous demonstrations showing long radiative lifetimes ranging from $\sim$20 ns to $\sim$ms, broad linewidths exceeding 600~\textmu eV, and a lack of reproducibility or tuning \cite{Zhao2021Site}. Moreover, no two-photon interference measurements have yet been reported for any MoTe$_2$-based emitter, and indistinguishability remains unverified for the entire TMD platform at telecom wavelengths. \par
In this work, we demonstrate a reproducible and deterministic approach to realizing high-quality and tunable quantum emitters in bilayer MoTe$_2$, exhibiting high single-photon purity and promising photon indistinguishability. By combining directional strain engineering with e-beam-induced defect activation, we generate stable and spectrally narrow emitters across multiple samples, operating in the 1090–1200 nm spectral range. These quantum emitters exhibit fast radiative decays of 130$-$450 ps, with a high degree of linear polarization (DOLP) up to $\sim$70$\%$, high single-photon purity ($g^{(2)}(0)< 0.1$), and resolution-limited emission linewidths of $\sim$150~\textmu eV. Quasi-resonant and p-shell excitations suppresses spectral diffusion and enhances purity (triggered $g^{(2)}(0) \sim 0.01$) through background emission reduction. Moreover, electrostatic biasing enables active tuning over a $\sim$3 meV range and reduces photon bunching.
A direct comparison of experimental emission linewidths ($W_\text{exp}$) and transform-limited linewidths ($W_\text{rad}$) reveals ultra-low ratios $R = W_\text{exp} / W_\text{rad}$ down to $R\sim55$, indicating significant suppression of charge noise.
Most notably, we observe two-photon interference with a Hong–Ou–Mandel visibility of $V_\text{HOM}\sim$~7.1$\%$, and up to $V_\text{HOM}\sim$~60$\%$ with post-selection (temporal filtering), representing the first demonstration of photon indistinguishability from a TMD-based quantum emitter in the near-infrared. These results establish MoTe$_2$ and TMD quantum emitters in general as a viable platform for tunable, low-noise single-photon sources capable of generating indistinguishable photons for photonic quantum technologies at telecom wavelengths.

\begin{figure}[ht!]
    \centering
    \includegraphics[width=0.95\columnwidth]{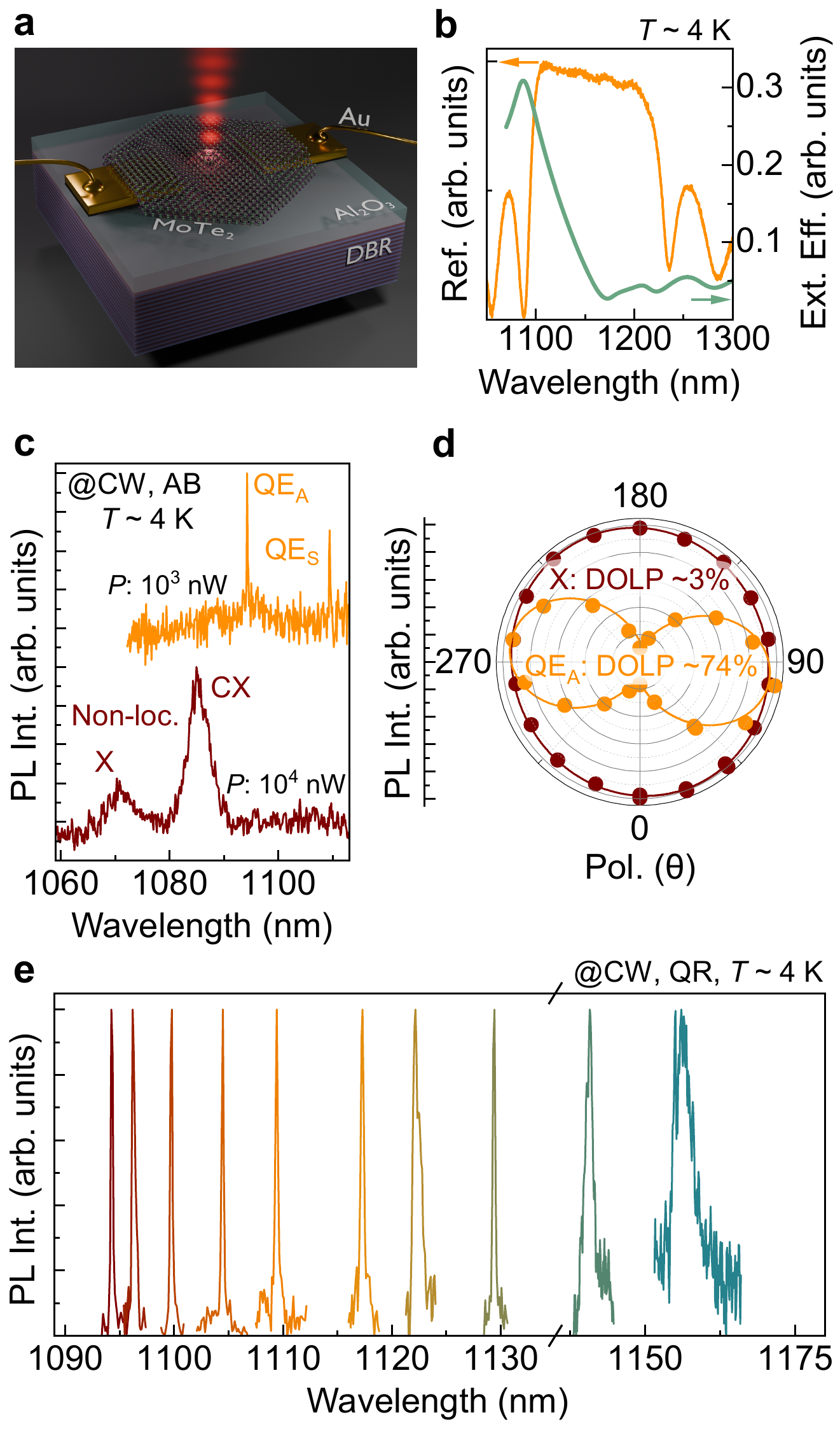}
    \caption{\textbf{Optical characterization of the sample (T $\sim 4$~K).} \textbf{a)} Schematic of the device architecture. A MoTe$_2$ bilayer transferred onto a DBR/Al$_2$O$_3$ substrate. Local strain induces a quantum emitter that produces single photons. Gold electrodes contact the flake, enabling an electric bias. \textbf{b)} Optical characterization of the DBR. The orange curve (left axis) shows near-unity reflectivity between 1100–1240~nm. The green curve (right axis) shows the calculated extraction efficiency, peaking around 1100~nm, well-matched to the MoTe$_2$ emission range. \textbf{c)} \textmu PL spectra under CW 650~nm (AB) excitation. Emission from a strained wrinkle region (orange) shows two sharp peaks (QE$_A$ and QE$_S$) attributed to localized quantum emitters. The unstrained region (dark red) displays broader emission assigned to the non-localized excitonic transition: exciton (X) and charged exciton (CX). \textbf{d)} Polarization profiles of QE$_A$ and delocalized excitonic transition X \textmu PL intensity. QE$_A$ shows a strong degree of linear polarization (DOLP $\sim74\%$), whereas X is unpolarized (DOLP $\sim3\%$). \textbf{e)} Representative normalized emission lines from multiple emitters across the sample, measured under quasi-resonant excitation (QR, 1080 nm).}
    \label{fig:one}
\end{figure}

\section{System Under Study}

To realize MoTe$_2$ quantum emitters (QE\textsubscript{A}), we begin by designing a highly reflective substrate consisting of a bottom distributed Bragg reflector (DBR) made of 20 pairs of GaAs (85.0 nm) and AlAs (99.7 nm) layers grown on a GaAs wafer, capped with an additional $\sim$100~nm thick Al$_2$O$_3$ layer (Fig.~\ref{fig:one}a). This substrate is specifically designed to maximize photon extraction efficiency \cite{Flatten2018Microcavity, Drawer2023Monolayer, Brotons2018Engineering} from MoTe$_2$ emitters fabricated on top (Fig.~\ref{fig:one}b), as supported by our optical simulations (see Methods and Supplementary Notes 1 and 2 for details on substrate preparation and extraction efficiency calculations). The measured reflectivity of the fabricated DBR substrate (orange line in Fig.~\ref{fig:one}b) exhibits high reflection (approaching unity) across a broad spectral range from 1100 nm to 1240 nm, well aligned with the emission spectrum of bilayer MoTe$_2$.
Gold electrodes are then fabricated via standard UV photolithography followed by metal deposition and lift-off. To implement directional strain engineering, star-shaped nanopillars are fabricated using a high-resolution negative resist (hydrogen silsesquioxane, or HSQ) via e-beam lithography. The resulting nanopillars are around 100 nm in height and feature a three-pointed star geometry, adapted from our previous work~\cite{Paralikis2025Tunable}.
Bilayer MoTe$_2$ flakes are mechanically exfoliated from bulk crystals (HQ-Graphene) using the standard scotch-tape method onto a polydimethylsiloxane (PDMS) stamp~\cite{Castellanos2014Deterministic}. The flakes are identified prior to transfer based on their photoluminescence (PL) spectra at room temperature (see Supplementary Note 1 and Methods for details).
Following identification, the selected bilayers are dry-transferred onto the substrate with pre-patterned nanopillars and electrodes (Fig.~\ref{fig:one}a).
During transfer, the bilayer conforms to the underlying three-pointed star nanopillars, forming nanowrinkles. These strain-induced wrinkle regions act as localized hosts for polarized quantum emitters via directional strain confinement, as demonstrated previously~\cite{paralikis2024tailoring}.
After the transfer, we performed optical imaging of the sample, finding promising strained regions which were then selectively irradiated by a focused e-beam to introduce atomic-scale defects, which activate the formation of localized quantum emitters~\cite{Linhart2019Localized, paralikis2024tailoring, Parto2021Defect}. \par
For initial optical characterization, we performed micro-photoluminescence (\textmu PL) measurements by placing the fabricated sample in a closed-cycle cryostat (attocube attoDRY800XS) operating at $T \sim 3.9$~K, equipped with a low-temperature objective ($\text{NA} = 0.80$) (further details in Methods). Under above-band (AB) excitation with a 650~nm laser (PicoQuant LDH-D-C-650), narrow PL emission lines (QE\textsubscript{A} and QE\textsubscript{S}) were obtained in strained wrinkle regions (orange line, Fig.~\ref{fig:one}c), consistent with localized states associated with single-photon emission in other TMD-based platforms~\cite{Linhart2019Localized, paralikis2024tailoring, Parto2021Defect}. In contrast, broader emissions (dark red line) are observed from the unstrained areas, attributed to delocalized exciton recombination~\cite{robert2016excitonic, moody2016exciton}. We further investigate the polarization properties of both the narrow emission lines and the broader excitonic peaks. Previous studies have shown that quantum emitters localized in nanowrinkles often exhibit a high degree of linear polarization (DOLP), typically aligned with the wrinkle direction~\cite{paralikis2024tailoring}. Consistent with these findings, QE\textsubscript{A} exhibits a pronounced polarization anisotropy, with a DOLP exceeding 70\% (Fig.~\ref{fig:one}d, orange line), in clear contrast to the unpolarized emission from the delocalized excitonic transitions (dark red line). Full characterization of delocalized excitonic transitions and other similar peaks is provided in Supplementary Note 3.
Lastly, we identified a range of narrow emission lines from multiple nanowrinkle regions across the sample, with wavelengths spanning from 1090~nm to 1170~nm (Fig.~\ref{fig:one}e). This spectral distribution is mostly associated with bilayer MoTe$_2$ emission and could potentially be extended to longer wavelengths toward the 2$^\text{nd}$ and 3$^\text{rd}$ telecom windows by exploiting multilayer MoTe$_2$. The observed narrow and linearly polarized emission lines from MoTe$_2$ nanowrinkles indicate a potential single-photon emission. To verify this and further investigate their single-photon properties, we perform systematic optical spectroscopy on the fabricated quantum emitters.

\begin{figure*}[t]
    \centering
    \includegraphics[width=0.9\textwidth]{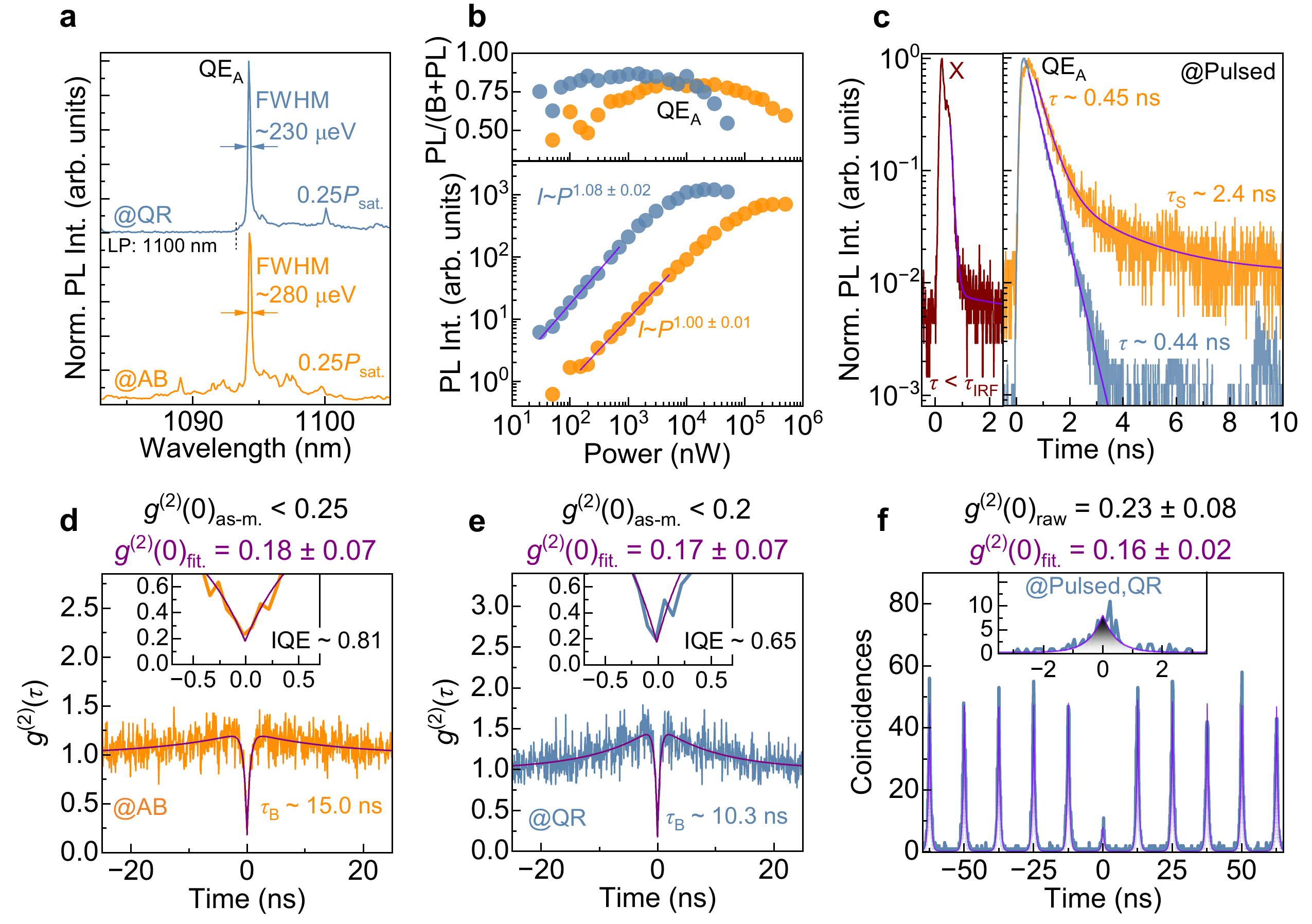}
    \caption{\textbf{QE$_{A}$ emission characterization and comparison between above-band (AB, 650~nm, orange) and quasi-resonant (QR, 1080~nm, blue) excitation schemes (T $\sim$~4~K).} \textbf{a)} Normalized \textmu PL spectra of QE$_A$ ($\sim$10~s integration time, $0.25 P_\text{sat}$, CW). QR excitation yields a lower background and a narrower linewidth (230~\textmu eV vs 280~\textmu eV). \textit{LP: 1100~nm} indicates the use of an additional tuned longpass filter (1100~nm) used for blocking the excitation laser. \textbf{b)} Top: PL-to-background ratio as a function of power (CW). QR excitation shows an improved PL-to-background ratio at low power and saturation, while AB shows an improved ratio for the saturation condition. Bottom: Saturation curves, both similar in shape (exponent between 1 and 1.1), but AB requires a higher power to saturate. \textbf{c)} TRPL traces under both excitation schemes (pulsed operation) and also from the delocalized exciton (dark red). The broad unlocalized excitonic transition decays rapidly ($\sim$80~ps), primarily limited by the instrument response function (IRF) of the excitation laser. AB excitation yields a biexponential decay ($\tau \sim 450$~ps, $\tau_S \sim 2.4$~ns), whereas QR yields a clean monoexponential decay with $\tau \sim 440$~ps for QE$_A$. \textbf{d)} Second-order autocorrelation measurement ($g^{(2)}(\tau)$) under CW AB excitation. Some bunching is observed, the fit yields a bunching time of $\tau_\text{B} \sim 15.0$~ns. The inset presents the antibunching dip around a limited time window and a fitted value of $g^{(2)}(0) = 0.18 \pm 0.07$. \textbf{e)} $g^{(2)}(\tau)$ under CW QR excitation. A more pronounced bunching is observed with a calculated $\tau_\text{B} \sim 10.3$~ns. The inset showing the antibunching dip reveals a fitted $g^{(2)}(0) = 0.17 \pm 0.07$, using a constrained fit (limited rise-time), avoiding underestimation -- see details in Supplementary Note~6. \textbf{f)} Pulsed QR $g^{(2)}(\tau)$ within a measurement window of $\pm$65~ns, with no observable bunching. The inset ($\pm$3.5~ns) presents the zero-delay peak with fitted value of $g^{(2)}(0) = 0.16 \pm 0.02$.}
    \label{fig:two}
\end{figure*}

\section{Single-photon emission}
Fig.~\ref{fig:two}a presents \textmu PL spectra ($T \sim 4$~K) taken from a representative quantum emitter (QE\textsubscript{A}), collected with the polarizer aligned to maximize the emission signal (corresponding to the polarization axis with high DOLP). To investigate how the excitation energy affects the emission properties, we collected \textmu PL spectra under two different excitation wavelengths of above-band 650~nm (orange, labeled AB) and quasi-resonant 1080~nm (blue, labeled QR). The quasi-resonant excitation results in a narrower emission line ($\sim$230~\textmu eV) with significantly suppressed background even under a longer integration time of 10~s, whereas the above-band excitation leads to broader emission ($\sim$280~\textmu eV) accompanied by strong, broad background emission. Moreover, when reducing the integration time from 10~s to 0.5~s under quasi-resonant excitation, the emission linewidth narrows further to the resolution-limited value of $\sim$150~\textmu eV (see more details in Supplementary Note 4). This narrowing under shorter integration time reflects slow ($>$0.5~s) spectral wandering due to charge fluctuations in the local environment, leading to inhomogeneous broadening over long integration windows. We additionally recorded a 150~s time trace of the emission, revealing line wandering bounded within $\sim$150~\textmu eV (Supplementary Note 4). Crucially, this slow wandering occurs on timescales orders of magnitude longer than the emitter lifetime ($\sim$440~ps) and the HOM delay between consecutive photons (12.5~ns), and therefore does not affect two-photon interference (see Section VI), in contrast to fast spectral diffusion on a few-ns timescale.  \par
To further evaluate the emitter response, we analyzed the PL intensity from QE\textsubscript{A} as a function of excitation power (Fig.~\ref{fig:two}b, bottom panel). For both excitation schemes, the integrated PL intensity scales with excitation power with an exponent of approximately 1–1.1, consistent with the behavior of a single excitonic transition. It also suggests that higher excitation powers are required to reach saturation under AB excitation compared to QR.
Additionally, the emission peak-to-background ratio (Fig.~\ref{fig:two}b, top panel) is consistently higher for QR excitation, especially below saturation, indicating reduced background contribution and enhanced spectral purity. Under saturation conditions, quasi-resonant excitation also yields $\sim$1.7 times higher emission intensity than the above-band excitation. This enhancement likely arises from more efficient population of the emitter under QR excitation, which directly addresses states close to the emitter resonance and thus avoids the carrier relaxation and non-radiative loss channels associated with above-band excitation through higher-energy bands. We note that the comparison of saturation intensity and PL-to-background ratio is intrinsically independent of the difference in saturation power between the two schemes, since those quantities are extracted at saturation. The difference in absolute saturation power may instead arise partly from the chromatic focal shift of the cryogenic objective between 650 and 1080 nm. \par
To investigate the recombination dynamics of these emitters, we performed time-resolved photoluminescence (TRPL) measurements under both pulsed AB and QR excitations with 80~MHz repetition rate. The TRPL trace of the delocalized exciton reveals an ultra-fast decay below 80~ps (Fig.~\ref{fig:two}c, dark red line), at the resolution limit of our setup, limited by the used pulsed laser, consistent with previous reports for TMDs-based delocalized excitonic transitions~ \cite{robert2016excitonic}. In contrast, QE\textsubscript{A} exhibits a substantially slower decay as well as excitation-dependent dynamics. Under both the AB (orange line) and QR (blue line) excitations, the fast component of the decay is similar, with $\tau \simeq$~440-450~ps, reflecting the radiative recombination of the localized emitter state. Under above-band excitation, however, an additional slower component with $\tau_S = 2.4$~ns is required to fit the data, indicating that a fraction of carriers reach the emitter through delayed feeding from nearby trap states populated by the high-energy above-band excitation pulse.
Under the quasi-resonant excitation, the decay is well described by a single exponential, consistent with the more direct population of the emitter discussed above. Additional TRPL measurements across the sample (Supplementary Note 5) reveal sub-nanosecond decays for all characterized emitters, with a mean lifetime of $\sim$263~ps and a range of 130-450~ps. \par
To quantify how close our emitters approach the transform-limited regime, we estimated the ratio $R = W_\text{exp} / W_\text{rad}$, where $W_\text{rad} = h/(2\pi \tau)$ is the lifetime-limited linewidth and $R=1$ corresponds to a Fourier-transform-limited emitter~\cite{Paralikis2025Tunable}. Using the fast decay component, we obtain $W_\text{rad} \sim 1.46$~\textmu{}eV under above-band excitation and $W_\text{rad} \sim 1.497$~\textmu{}eV under the quasi-resonant excitation, yielding $R_{AB} = 192$ ($W_\text{exp, 10s} = 280$~\textmu{}eV) and $R_{QR} = 154$ ($W_\text{exp, 10s} = 230$~\textmu{}eV). Reducing the integration time to 0.5~s under QR excitation, where charge noise-induced wandering is less present, gives $W_\text{exp, 0.5s} = 150$~\textmu{}eV and $R_{QR} = 100$, which, to our knowledge, is among the lowest reported values for non-cavity-enhanced TMD-based quantum emitters~\cite{Paralikis2025Tunable}. The remaining broadening reflects additional dephasing channels, while the comparatively short lifetimes of our emitters are consistent with previous studies on another molybdenum-based TMD (MoSe$_2$) quantum emitter~\cite{Yu2021Site}. \par
To verify the single-photon nature of the emission and quantify its purity, we performed second-order autocorrelation measurements under CW, under both above-band (Fig.~\ref{fig:two}d) and quasi-resonant (Fig.~\ref{fig:two}e) excitations. In both cases, the measured raw $g^{(2)}(0)$ values below 0.25 indicate strong photon antibunching. Fitting the data yields $g^{(2)}(0) = 0.18 \pm 0.07$ for above-band and $0.17 \pm 0.07$ for quasi-resonant (details about fitting, avoiding underestimation provided in Supplementary Note~6), confirming the single-photon nature of the emission from QE$_A$. While the quasi-resonant fit is centered at a lower value, the two results are within their respective uncertainties, not signifying any quantitative improvement in single-photon purity from quasi-resonant excitation alone -- a clear advantage requires the optimized excitation conditions demonstrated in Section~V.
Beyond purity, the $g^{(2)}(\tau)$ measurements also offer insight into the emitter internal dynamics. Under both CW excitation schemes, we observed bunching around the central antibunching dip, typically attributed to blinking behavior on timescales within a measurement window (here $\pm$1~\textmu s). Including this additional dynamics in the fit, we extracted bunching times of $\sim$15.0~ns for above-band and $\sim$10.3~ns for quasi-resonant excitation. These values, along with the extracted on/off decay times ($\tau_\text{on}/\tau_\text{off}=78$~ns$/$18 ns for above-band and 30 ns$/$16 ns for quasi-resonant), reflect the influence of non-radiative processes contributing to the bunching dynamics. 
We attribute the observed blinking to several possible mechanisms. A primary contributor is spectral diffusion, where charge fluctuations in the local environment cause emission energy shifts. If spectral filtering is narrow, this can lead to intermittent signal detection. Additionally, nearby defect states may transiently trap charge carriers, reducing the excitation efficiency and thus affecting the emitter recombination dynamics. These processes can also modulate the relative populations of excitonic species, resulting in temporal switching between neutral and charged exciton configurations.
Despite their distinct origins, all these mechanisms reduce the internal quantum efficiency (IQE) by rendering the emitter inactive during “off” intervals. From the fitted bunching amplitudes in the autocorrelation data, we estimated IQEs of $\sim$0.81 and $\sim$0.65 under CW above-band and quasi-resonant excitations, respectively. The higher IQE, under the broader, higher-energy carrier generation (AB excitation) reflects a more efficient saturation of these slow trap states -- distinct from the fast trap-mediated feeding processes visible in TRPL. Importantly, the IQE difference between AB and QR does not directly translate to a difference in single-photon purity, since IQE captures the fraction of time the emitter is active, while $g^{(2)}(0)$ captures the multi-photon emission probability when it is active. Furthermore, power-dependent second-order autocorrelation measurements under CW quasi-resonant excitation (Supplementary Note 4) reveal reduced bunching - and thus higher IQE - at elevated excitation powers. This behavior is likely linked to saturation of nearby charge traps, providing additional evidence for the key role of charge fluctuations and their dependence on excitation conditions. \par
In addition to CW excitation, we performed second-order correlation measurements under quasi-resonant pulsed excitation (Fig.~\ref{fig:two}f). The raw data analysis yields a $g^{(2)}(0) = 0.23~\pm~0.08$, based on comparing the zero-delay peak area to the average side peaks. A fit to the autocorrelation histogram yields a slightly lower value of $g^{(2)}(0) = 0.16~\pm~0.02$. To understand the origin of these nonzero values, we compared the data fitted values (see Fig.~\ref{fig:two} and Supplementary Note~6) to those calculated for an ideal single-photon emitter contaminated by the measured background~\cite{Brouri2000Photon}, with parameters extracted from the integrated signal-to-background ratio in the PL spectra. For CW above-band excitation, this calculation yields $g_{BG-limited}^{(2)}(0) \sim 0.19$, while for CW and pulsed quasi-resonant excitations the corresponding values are $\sim$0.06 and $\sim$0.12, respectively. These estimates are consistent with our data, confirming that background emission plays a dominant role and QE$_A$ exhibits near-unity single-photon purity. However, the results presented here, especially the comparison between single-photon emission under AB and QR excitation, are still limited, hindering a clear advantage of QR excitation for background emission suppression. To better address this issue of increasing single-photon purity under optimal excitation, we also performed p-shell excitation, significantly improving single-photon purity (Section V).  \par
The pulsed $g^{(2)}(\tau)$ histogram (Fig.~\ref{fig:two}f), acquired during the same cooldown as the CW data, shows no observable bunching. However, a second dataset from a different cooldown (Supplementary Note 7) reveals clear bunching under pulsed excitation. This suggests the presence of blinking dynamics and reduced IQE, which were likely masked in Fig.~\ref{fig:two}f due to limited coincidence statistics. Differences in the local charge environment between cooldowns may also contribute to the observed variation in bunching amplitude. Despite these variations, all measurements confirm the single-photon nature of QE$_A$'s emission. The emitter displays highly pure and strongly polarized quantum light, with a sub-nanosecond lifetime, marking it as a high-quality single-photon source. \par
To demonstrate the reproducibility of our fabrication method, we also characterized additional quantum emitters from other wrinkled sites within the same sample (Supplementary Note 8) and in other samples fabricated with the same method (Supplementary Note 9). As summarized in Supplementary Notes 5, 8, and 9, these emitters exhibit similar optical properties, including reproducible narrow emission lines, high DOLP, short lifetimes, and high-purity single-photon emission, further confirming the robustness and scalability of our fabrication approach. Besides observing QEs for naturally occurring defects within the strained MoTe$_2$ area, we also investigated an additional sample to verify the yield of our fabrication method (Supplementary Note 9). From this characterization, we observed efficient (50\% of the available sites) generation of quantum emitters when both strain and defect engineering are performed (Supplementary Note 9), underscoring the importance of deterministic e-beam-based defect formation for a higher QEs fabrication yield. While the presented results demonstrate the realization of single-photon emitters, their linewidths are limited by local charge fluctuations and remain a dominant source of blinking and spectral instability. In the following section, we demonstrate how an external electric field can be used to control the local charge environment, enable spectral tuning, and enhance emission stability.

\begin{figure*}[]
    \centering
    \includegraphics[width=0.8\textwidth]{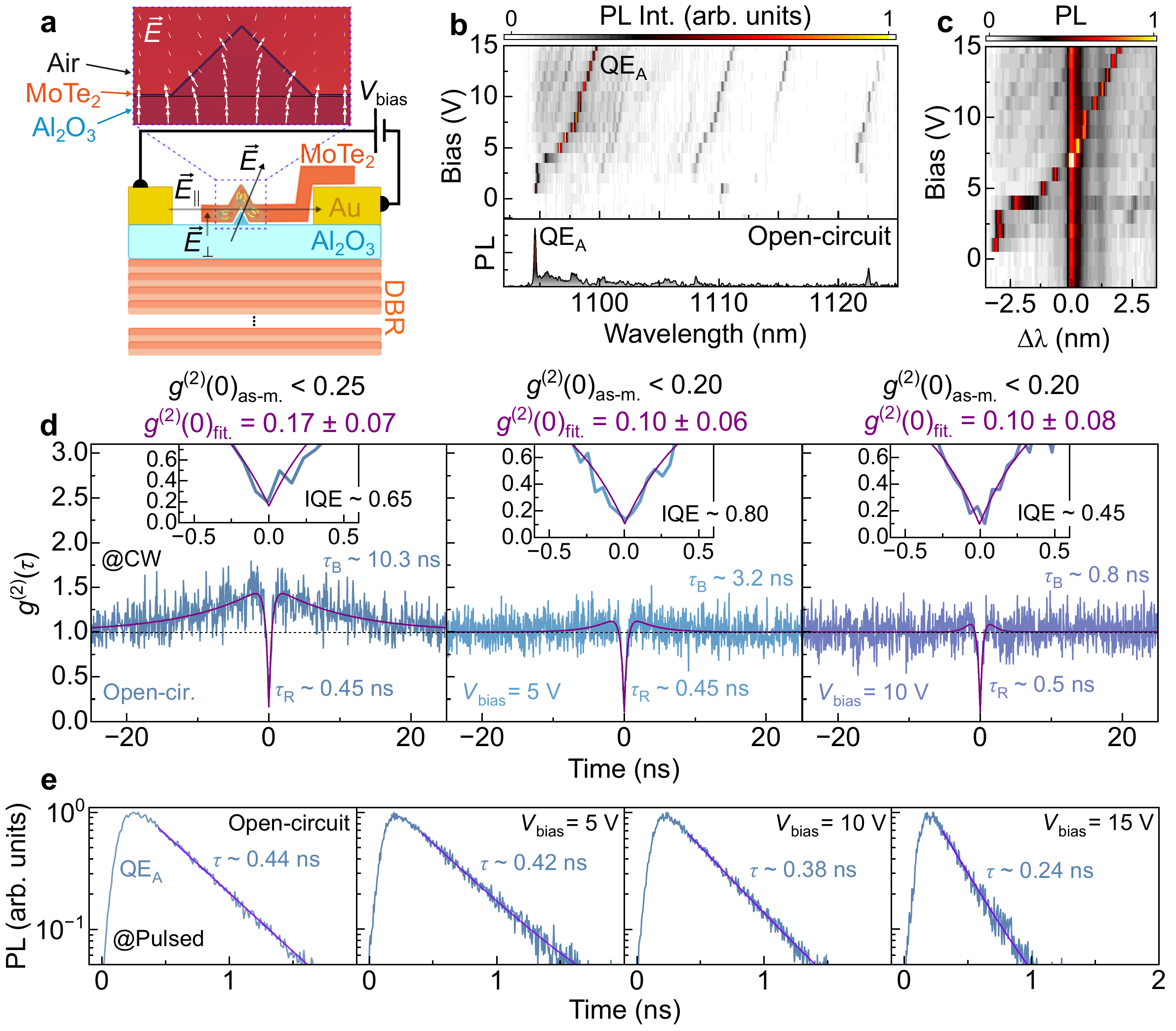}
    \caption{\textbf{Effects of biasing on the QE$_A$ optical properties under CW and pulsed quasi-resonant (QR) excitation at T $\sim$ 4~K.} \textbf{a)} Cross-sectional schematic of the biased device. The bilayer MoTe$_2$ flake sits on HSQ nanopillars (for strain engineering) and the DBR/Al$_2$O$_3$ substrate, in direct contact with one gold (Au) electrode (right). A second Au electrode positioned $\sim$20~µm laterally on the substrate (left) serves as the counter-electrode; the MoTe$_2$ flake ends before reaching this electrode. Applying a bias voltage $V_{\text{bias}}$ between the two electrodes produces an electric field with both in-plane ($\vec{E}_\parallel$) and out-of-plane ($\vec{E}_\perp$) components around the emitter, rather than a purely in-plane field. This is supported by the inset showing a zoom-in of the electric field simulation around the strained region of the flake under a $+$15~V (see Supplementary Note 10 for details). \textbf{b)} The bottom panel shows a \textmu PL spectrum (CW) of QE$_A$ at open-circuit condition, revealing additional emission from other lines. The top panel depicts the bias-dependent evolution of the same spectral range from $-$2~V to $+$15~V. Emission quenches under negative bias, while increasing bias results in intensity enhancement and a clear redshift of QE$_A$ and nearby peaks. \textbf{c)} Overlay of the bias-tuned QE$_A$ PL map with a reference emitter of similar emission energy at open-circuit. As the bias increases, QE$_A$ tunes across the reference peak, demonstrating the possibility of bias-driven spectral matching between distinct emitters. \textbf{d)} CW QR second-order autocorrelation measurements of QE$_A$ at open circuit, $+5$~V, and $+10$~V. As bias increases, bunching becomes less pronounced, and the extracted bunching times decrease from $\tau_\text{B} \sim 10.3$~ns to $\sim$3.2~ns to $\sim$0.8~ns. Insets show antibunching dips, with the fitted $g^{(2)}(0)$ improving from $0.17 \pm 0.07$ (0~V) to $0.10 \pm 0.06$ ($+5$~V) and $0.10 \pm 0.08$ ($+10$~V). Constrained fits with limited rise-time ($\tau_\text{R}$) are used to avoid $g^{(2)}(0)$ value underestimation (see details in Supplementary Note~6). \textbf{e)} Normalized TRPL measurements of QE$_A$ (pulsed excitation) under four bias configurations (open circuit, $+5$~V, $+10$~V, $+15$~V). As the bias increases, we observed decreasing decay times, with the monoexponential fits yielding $\sim$440~ps, $\sim$420~ps, $\sim$380~ps, and $\sim$240~ps for 0~V, +5~V, +10~V, and +15~V, respectively.}
    \label{fig:three}
\end{figure*}

\section{Electrical tuning}

To investigate how external electric fields can be used to control both the emission energy and the local charge environment of the quantum emitters, we now study their optical properties under applied electrical bias. Electrical biasing affects the optical properties of the emitters in two complementary ways: (i) a tuning of the emission line, enabling spectral shift of an individual emitter, and (ii) modulation of the local charge configuration, affecting blinking, lifetime, and the linewidth-to-lifetime ratio. 
Fig.~\ref{fig:three}a illustrates the cross-section of an electrically contacted device. The unencapsulated bilayer MoTe$_2$ flake sits on the DBR/Al$_2$O$_3$ substrate with HSQ nanopillars for strain engineering, and is in direct contact with one pre-patterned gold electrode (right). The flake ends before reaching a second gold electrode (left) positioned $\sim$20~µm laterally on the substrate, which serves as the counter-electrode. A bias voltage applied between these two electrodes establishes an electric field with both in-plane ($\vec{E}_\parallel$) and out-of-plane ($\vec{E}_\perp$) components across the strain-engineered region of the flake, rather than a purely in-plane field, following the laterally separated two-electrode biasing method we previously demonstrated for other TMD emitters~\cite{Paralikis2025Tunable}. Because the MoTe$_2$ flake is in direct contact with the one electrode only, the applied bias directly modulates the electrostatic potential of the flake, while the field is established between the two electrodes through the Al$_2$O$_3$ substrate underneath the flake, the MoTe$_2$ bilayer itself, and the gap region between the flake edge and the counter-electrode. Numerical simulations of the electric field (Fig.~\ref{fig:three}a, inset) confirm the coexistence of both field components (in-plane along the flake and out-of-plane through the substrate) near the emitter under applied bias for this geometry (see Supplementary Note~10 for details). Due to the intrinsic complexity and variability of strained TMD emitters, the exact configuration between the emitter dipole and the resulting electric field cannot be determined precisely from the present data. We therefore attribute the observed electrical tuning to bias-induced changes in the local electrostatic environment rather than to a conventional Stark effect with a uniquely defined field orientation.
Fig.~\ref{fig:three}b presents the bias-dependent \textmu PL response of the exemplary emitter QE$_A$ (bottom panel) under CW quasi-resonant excitation. As $V_\text{bias}$ increases from $-2$~V to $+15$~V, the main emission line exhibits a clear redshift starting near $V_\text{bias} = +4$~V. The total tuning range exceeds 3 nm ($\sim$3 meV), enabling spectral alignment of an emission line to another individual emission line as demonstrated in Fig.~\ref{fig:three}c. This tunability is a prerequisite for remote quantum interference between separate sources. Interestingly, the electrical tuning is strongly polarity-dependent; positive bias enhances PL intensity and induces tunability, while negative bias suppresses emission without a significant spectral shift. Occasionally, abrupt step-like switching between two emission lines is observed near $V_\text{bias} = +4$~V, likely caused by changes in the local charge environment and formation of different excitonic complexes (see more detailed analysis in Supplementary Note 10). \par
The experimental linewidth of QE\textsubscript{A} under bias remains stable at $W_\text{exp} \sim 150$~\textmu{}eV (0.5~s integration time) across all bias conditions (Supplementary Note 10), close to our setup's resolution limit ($\sim$140-160~\textmu eV). The fast ($<$0.5~s) spectral diffusion responsible for this $\sim$150~\textmu{}eV broadening -- which partially sets the dephasing relevant for two-photon interference (see Section~VI) -- is therefore unaffected by the applied bias. On longer timescales, however, we observe a modest increase in spectral wandering under bias, from $\sim$100~\textmu eV at open-circuit to $\sim$250~\textmu eV at higher bias voltages (Supplementary Note 10), likely arising from slow reconfiguration of charge traps in response to the applied field. As we discuss in Section~VI, this slow wandering occurs orders of magnitude more slowly than the radiative lifetime and the delay between consecutive photons (12.5~ns), and therefore does not affect single-photon properties. \par
Biasing also influences recombination dynamics and blinking dynamics. Second-order autocorrelation measurements in Fig.~\ref{fig:three}d show a $g^{(2)}(0)$ value of $0.10 \pm 0.06$ at $+5$~V and $0.10 \pm 0.08$ at $+10$~V (see details about fitting, avoiding underestimation in Supplementary Note~6). While the limited photon statistics do not robustly support a bias-induced enhancement of single-photon purity, the bunching contribution is most likely suppressed under bias. The weaker bunching contribution under bias complicates fitting and parameter estimation, as detailed in Supplementary Note~6. Nevertheless, bunching times decrease from $\tau_\text{B} \sim 10.6$~ns at open-circuit to $\sim$3.2 ns at $+5$~V and $\sim$0.8 ns at $+10$~V, and the reduced bunching amplitude corresponds to an increase in IQE from $\sim$0.65 at open-circuit to $\sim$0.80 at $+5$~V (fits at $+10$~V are photon-statistics-limited). Both effects support a downward trend compared to open-circuit conditions, demonstrating suppression of blinking under bias~\cite{Cai2024Charge}.
Time-resolved PL measurements (Fig.~\ref{fig:three}e) reveal a pronounced reduction in lifetime from $\sim$440 ps at $0~$V to $\sim$240 ps at $+15~$V, a more than 45$\%$ decrease consistent with prior observations~\cite{Paralikis2025Tunable}. Since the measured linewidths remain resolution-limited ($\sim$150~\textmu eV), the corresponding linewidth-to-lifetime ratio R decreases from $R_\text{0V}=100$ to $R_\text{15V}=55$, approaching the transform-limited regime. To our knowledge, this represents one of the lowest reported $R$ values for any non-cavity-enhanced TMD-based emitters~\cite{Paralikis2025Tunable}.
The observed reduction in lifetime can be attributed to two complementary effects that contribute at different bias voltages. Below approximately $+8$~V, where the PL intensity increases with bias, the lifetime reduction is consistent with a bias-induced enhancement of radiative recombination~\cite{Polland1985Lifetime}. The microscopic origin — potentially involving field-induced modulation of the excitonic wavefunctions or of the local charge environment around the emitter — cannot be uniquely determined from our current data, but similar bias-induced enhancement of radiative rates has been reported for other TMD-based~\cite{Paralikis2025Tunable} and QD-based emitters~\cite{Undeutsch2025Electrical}. Above $+10$~V, where the PL intensity decreases, additional non-radiative pathways likely become significant, plausibly through thermal population of dark or higher-energy excitonic states enabled by increased phonon population from leakage-induced heating~\cite{Lehner2023Beyond, Smith2005Voltage, Stevens2022Enhancing}. Both mechanisms can shorten the observed decay time, and the relative contribution of each depends on the applied bias voltage.
Overall, electrical biasing provides robust control over both the emission energy and radiative dynamics of MoTe$_2$ quantum emitters. Taken together, the applied bias provides two complementary forms of control over MoTe$_2$ quantum emitters. The electrical tuning enables continuous spectral shift over $\sim$3~meV, sufficient to bring independent emitters into spectral resonance. Modulation of the local charge environment simultaneously suppresses long-timescale blinking and reduces the radiative lifetime, leading to a near-halving of the linewidth-to-lifetime ratio toward the transform limit, with only a modest increase in slow spectral wandering that does not affect two-photon interference. While the bias study addresses charge-noise-driven blinking and lifetime, two further avenues remain to be explored. Namely, dielectric environment control through full hBN encapsulation, and optimized excitation through higher-energy resonances. We address both in the following section.

\begin{figure*}[t]
    \centering
    \includegraphics[width=0.85\textwidth]{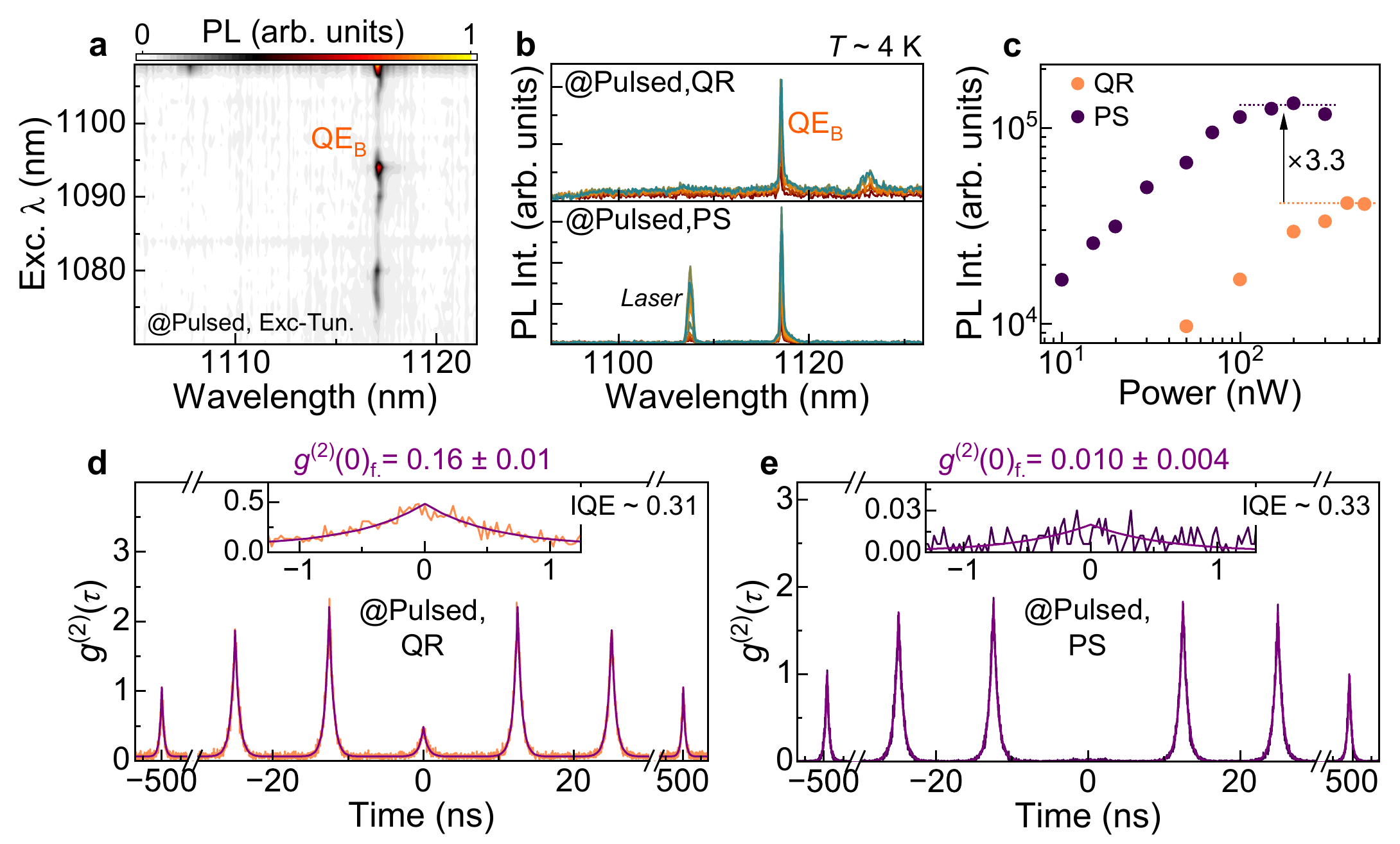}
    \caption{\textbf{Impact of advanced excitation on hBN-encapsulated quantum emitter (QE$_\text{B}$) at T $\sim$ 4~K.} \textbf{a)} Photoluminescence excitation (PLE) map for QE$_\text{B}$ recorded by tuning the pulsed excitation laser (see details in Methods), showing several higher energy states for efficient stimulation of investigated emitters. \textbf{b)} \textmu PL spectra recorded under several excitation powers using both pulsed quasi-resonant (QR) in top-panel and pulsed p-shell (PS) excitation (excitation wavelength~$\sim1108$~nm) in bottom-panel ($\sim$10~s integration time). \textbf{c)} Saturation curves for QE$_\text{B}$ under both QR and PS conditions extracted from \textbf{b)}, showing similar exponent in both cases but significantly improved saturation emission intensity under PS conditions. \textbf{d)} Second-order correlation measurement $g^{(2)}(\tau)$ under pulsed QR excitation. The data show clear bunching, with an extracted internal quantum efficiency (IQE) of $\sim$31\%. The inset displays the central region ($\pm$1.25~ns), where the fit yields $g^{(2)}(0) = 0.16~\pm~0.01$. \textbf{e)} $g^{(2)}(\tau)$ under pulsed PS excitation (excitation wavelength $\sim1108$~nm). The data show clear bunching, with an extracted IQE of $\sim$33\%. The inset displays the central region ($\pm$1.25~ns), where the fit yields $g^{(2)}(0) = 0.010~\pm~0.004$.}
    \label{fig:four}
\end{figure*}

\section{hBN-encapsulation and optimized excitation}

Building on the bias study, we now investigate two further routes for improving emitter performance: full hexagonal boron nitride (hBN) encapsulation, and excitation through higher-energy resonances identified by photoluminescence excitation spectroscopy (PLE). Encapsulation in hBN has been widely used to stabilize TMD-based emitters by providing a clean dielectric environment and reducing surface-induced charge noise~\cite{klein2019site, chakraborty2019electrical, Paralikis2025Tunable, He2016Cascade, Iff2017Substrate, Kutrowska2022Exploring}. In our system, encapsulation also modifies the strain landscape, replacing the directional nanowrinkles with a more bubble-like morphology (Supplementary Note 1).
A careful characterization of multiple hBN-encapsulated emitters reveals optical properties broadly similar to the non-encapsulated case (Supplementary Notes 5, 11, and 12). We identify isolated narrow lines with comparable lifetimes and single-photon purity under quasi-resonant excitation. However, three quantitative differences emerge. First, the bubble-like morphology produces weaker confinement asymmetry, reflected in lower DOLP than in non-encapsulated QEs (Supplementary Notes 5, 11, and 12). Second, most encapsulated lines reveal two features with orthogonal polarizations, indicating fine-structure splitting (FSS) of the localized exciton state, in contrast to non-encapsulated emitters subjected to stronger wrinkle-induced strain, which exhibit larger FSS and more imbalanced emission intensities (Supplementary Note 13). Third, encapsulated QEs show slightly longer lifetimes than non-encapsulated emitters (Supplementary Note 5), consistent with weaker strain confinement and reduced oscillator strength, likely arising from the rigidity of hBN promoting bubble formation rather than tight wrinkle~\cite{androulidakis2018tailoring, Daveau2020Spectral}. \par
While hBN encapsulation does not reduce charge noise, it provides a clear advantage in long-term stability against ambient exposure. Encapsulated emitters retain their high-performance optical properties across multiple cooldowns over an extended period (Supplementary Notes 14 and 15), whereas non-encapsulated emitters survive long-term ambient exposure but exhibit degraded stability that limits advanced optical characterization (Supplementary Note 16). Encapsulated emitters also more frequently emit at longer wavelengths above 1110~nm, well-detuned from the non-localized bilayer excitonic transition. We note that QR excitation at 1080~nm couples through this non-localized excitonic state and can therefore be applied generically across many emitters (Supplementary Note 17), but it does not directly address the localized emitter or its higher-energy resonances. The larger detuning of encapsulated emitters at above 1110~nm instead enables direct addressing of higher-energy localized states -- a capability that motivates the optimized excitation studies described below. \par
To identify these resonances, we performed photoluminescence excitation spectroscopy on QE$_\text{B}$ by tuning the pulsed excitation wavelength. The PLE map (Fig.~\ref{fig:four}a) reveals several resonances corresponding to higher-energy states of the localized emitter, with the strongest response at the smallest detuning from the emission line. Excitation at this resonance ($\sim$1108~nm) -- which we identify as a p-shell-like state -- produces a striking improvement in emitter performance compared to QR excitation. Fig.~\ref{fig:four}b shows that p-shell (PS) excitation strongly suppresses the background emission visible under QR excitation, with near-complete background extinction. The saturation curves in Fig.~\ref{fig:four}c reveal a $3.3\times$ improvement in saturated brightness under PS excitation. Comparable performance is obtained at one further resonance identified in the PLE map (Supplementary Note 14), confirming that the improvement is a general feature of higher-energy excitation rather than specific to one resonance. \par
The most striking effect of PS excitation is on single-photon purity (Figs.~\ref{fig:four}d,e). The strongly suppressed background under PS excitation enables measurement of $g^{(2)}(\tau)$ with minimal spectral filtering, more specifically, a tunable longpass filter and a tunable bandpass filter ($\sim$10~nm spectral width) centered near the emission wavelength. The fitted pulsed $g^{(2)}(0)$ improves from $0.16\pm0.02$ under QR excitation to $0.010\pm0.004$ under PS excitation, a more than tenfold reduction driven entirely by the suppression of background emission. Both measurements reveal substantial bunching, with a limited IQE ($\sim31-33\%$), likely arising from shallower confinement in the bubble strain landscape (Supplementary Notes 11, 12, and 15). To verify the robustness of the scheme, we measured the pulsed $g^{(2)}(0)$ under PS excitation on two additional emitters (Supplementary Note 15), obtaining consistently low values and confirming that PS excitation reliably yields high single-photon purity across hBN-encapsulated MoTe$_{2}$. Having characterized our emitters under no encapsulation, hBN encapsulation, electrically biased, and p-shell-excited conditions, we now address the central remaining question: photon indistinguishability. Achieving high indistinguishability is the principal outstanding challenge for TMD-based quantum emitters and a key benchmark for their use in scalable quantum photonic technologies. \par

\begin{figure*}[t]
    \centering
    \includegraphics[width=0.99\textwidth]{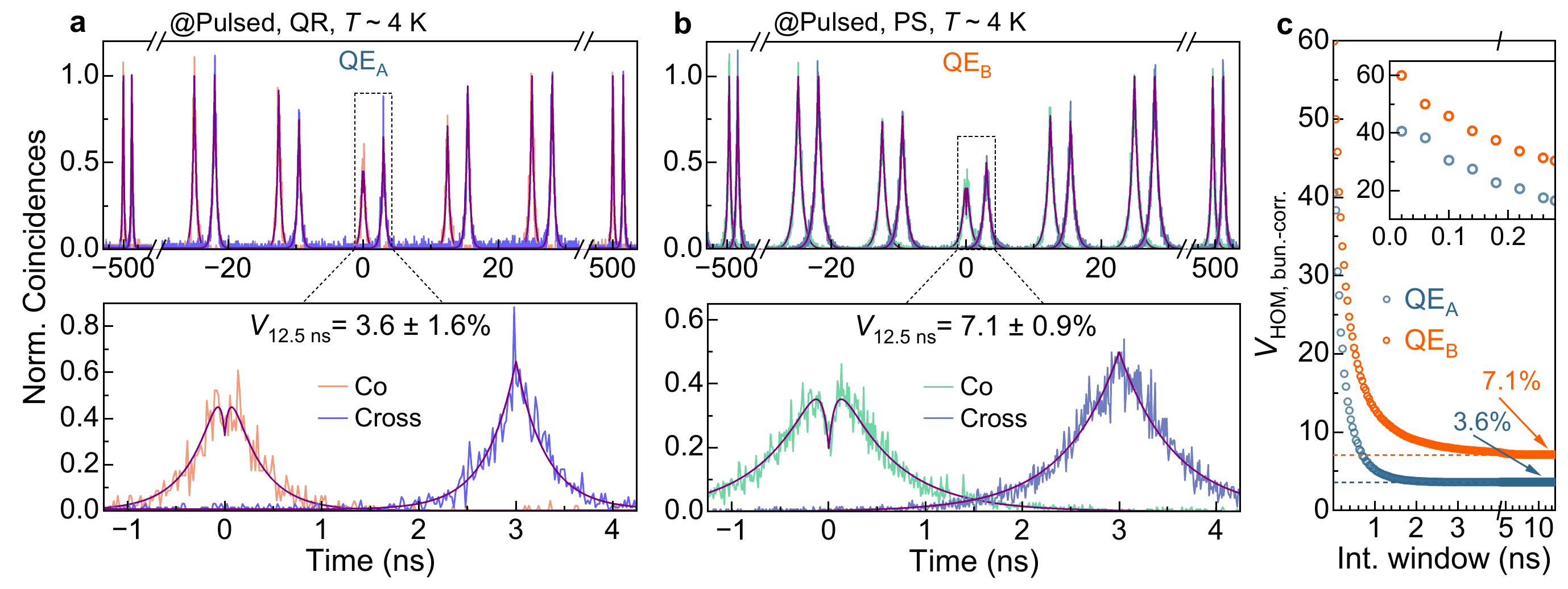}
    \caption{\textbf{Two-photon interference HOM measurement for QE$_A$ under pulsed QR and QE$_B$ under pulsed PS excitation at $\sim$4~K.} The bunching-corrected HOM measurements under both polarization configurations for \textbf{a)} QE$_A$ (co-polarized configuration in red, cross-polarized configuration in blue) and for \textbf{b)} QE$_B$ (co-polarized configuration in green, cross-polarized configuration in blue). For clarity, the cross-polarized data are slightly time-shifted to enhance visual distinction. The data are presented over a wide time range ($\pm$507~ns) out of the bunching region. Purple lines indicate the fits for both configurations. The lower panels present a magnified plot of the zero-delay peak, showing a clear difference between co- and cross-polarized configurations. A comparison of the integrated coincidences in a 12.5 ns time integration window for both configurations indicates \textbf{a)} $V_{\text{12.5 ns}}$ $=$ 3.6 $\pm$ 1.6~$\%$ for QE$_A$ and \textbf{b)} $V_{\text{12.5 ns}}$ $=$ 7.1 $\pm$ 0.9~$\%$ for QE$_B$. \textbf{c)} Extracted visibility of two-photon interference for different time integration windows for QE$_A$ (blue points) and QE$_B$ (orange points). The data show visibility values higher than $\sim$20$\%$ (QE$_A$) and even higher than $\sim$30$\%$ (QE$_B$) for reduced integration windows below 200 ps, while longer time windows feature a reduction in visibility, which saturates at longer time scales.
    }
    \label{fig:five}
\end{figure*}

\section{Photon indistinguishability}

To probe the indistinguishability of the MoTe$_{2}$ quantum emitters, we performed Hong-Ou-Mandel (HOM) two-photon interference measurements on QE$_A$ and QE$_B$, under pulsed quasi-resonant and p-shell excitation, respectively, which allows us to assess the impact of optimized excitation on photon indistinguishability. Both emitters retained their single-photon emission across multiple cooldowns. Under the HOM measurement settings, QE$_B$ maintained its high single-photon purity demonstrated in Section~V ($g^{(2)}(0)$ = $0.010\pm0.004$, Fig.~\ref{fig:four}e), while QE$_A$, operated at a higher excitation power to improve count rates in the HOM measurement, exhibited an increased $g^{(2)}(0)$ value below 0.35, with some additional bunching and re-excitation signatures (Supplementary Note 7). The HOM interferometer used a free-space input beamsplitter, a 12.5~ns fiber delay line matched to the laser repetition rate, and a fiber-based output beamsplitter, with the polarization in one arm rotated by a half-wave plate to switch between co-polarized and cross-polarized configurations~\cite{Madigawa2025Deterministic}. We fitted the raw coincidence histograms using a standard pulsed-HOM model that includes a bunching component to account for residual photon correlations at longer timescales~\cite{Vajner2024OnDemand, Madigawa2025Deterministic}. The fitted bunching parameters were then used to correct the raw data for a fair comparison between co- and cross-polarized configurations, after which we re-fitted the bunching-corrected data with the same model excluding the bunching component. \par
Figures.~\ref{fig:five}a,b present the bunching-corrected coincidence histograms over a $\pm507$~ns time window, with the cross-polarized data slightly time-shifted for clarity. The extended window includes peaks far from zero delay that are unaffected by bunching, providing a reliable normalization reference. The slight asymmetry of the side peaks at $\pm12.5$~ns for QE$_A$ arises from a modest imbalance between the two arms of the output beamsplitter, whose specified optimal wavelength range (930~$\pm$~100 nm) does not fully cover the emission of our QEs at $\sim$1100~nm. The bottom panels of Figs.~\ref{fig:five}a,b show a magnified view of the zero-delay region, where the bunching-corrected peak areas integrated over the full 12.5 ns repetition window, yield an HOM visibility of $V_{\text{12.5 ns}} =3.6 \pm 1.6\%$ for QE$_A$ and $V_{\text{12.5 ns}} = 7.1 \pm 0.9\%$ for QE$_B$. Temporal post-selection produces substantially higher visibilities, exceeding $>$20$\%$ for QE$_A$ and $>$30$\%$ for QE$_B$ when restricting the integration window to 200~ps, and reaching up to $\sim$40\% (QE$_A$) and $\sim$60\% (QE$_B$) under further temporal filtering toward the smallest accessible windows (Fig.~\ref{fig:five}c). For longer integration windows, the visibility decreases and saturates at the full-window value, consistent with the dephasing-limited nature of the interference. To establish the reproducibility of these results, we performed three independent HOM measurements on QE$_B$ on separate days under nominally identical conditions, all yielding visibilities consistent with the value reported above (Supplementary Note 18). We additionally measured a second hBN-encapsulated emitter under PS excitation, obtaining $V_{\text{12.5 ns}} =4.2 \pm 0.8\%$ (Supplementary Note 18), confirming that two-photon interference is a robust feature of MoTe$_2$ quantum emitters under optimized excitation. \par
From the bunching-corrected HOM fits, we also estimate the emitter’s coherence time to be $\tau_c =62 \pm 15$~ps for QE$_A$ and $\tau_c =81 \pm 10$~ps for QE$_B$. The fit uncertainty for QE$_A$ is larger because of its lower visibility and signal-to-noise ratio, while the cleaner dip in QE$_B$ -- combined with our setup's $\sim$50~ps timing resolution -- yields a more precise estimate. These short coherence times, compared to the radiative lifetimes ($\tau$~$\sim$~240-440~ps for QE$_A$ and $\tau$~$\sim$~500~ps for QE$_B$), account for the modest visibilities observed within the 12.5~ns window.
Moreover, it is important to distinguish between the slow spectral wandering observed under second-scale integration ($\sim$150-200 \textmu eV) and the fast dephasing relevant for two-photon interference, which operates on the $\sim$10 ns timescale set by our HOM delay line. Using the standard relation $V = \tau_c / (2\tau)$ for a pulsed HOM experiment limited by pure dephasing~\cite{Bylander2003Interference,Santori2004Single}, the coherence times extracted above yield expected visibilities of $\sim$7-13\% (QE$_A$, using $\tau \sim 240-440$~ps) and $\sim$8\% (QE$_B$, using $\tau \sim 500$~ps). For QE$_B$, this is in excellent agreement with the measured $V_{\text{12.5 ns}} = 7.1 \pm 0.9\%$, while for QE$_A$ the measured $V_{\text{12.5 ns}} = 3.6 \pm 1.6\%$ is further reduced by the elevated $g^{(2)}(0) \sim 0.35$ under the high-power conditions used in the HOM experiment. This consistency demonstrates that our visibilities align with the standard theoretical framework, and that the long-timescale spectral wandering observed under long-integration spectroscopy does not directly limit two-photon interference.
Taken together, our results demonstrate two-photon interference with a visibility of  $\sim$3-7\% within the 12.5~ns integration window, rising to $>$40-60\% under sub-200~ps post-selection. To the best of our knowledge, this represents the highest reported indistinguishability for any TMD-based quantum emitter, and the first such demonstration for MoTe$_2$ emitters.

\section{Discussion}
Our results demonstrate the reproducible generation and comprehensive optical characterization of single-photon emitters in bilayer MoTe$_2$ within the near-infrared spectral range (1090–1200 nm). Combining strain engineering via nanopillars with controlled e-beam-induced defect activation, we established a robust fabrication process yielding MoTe$_2$ quantum emitters with narrow linewidths down to $\sim$150 \textmu eV, radiative lifetime as short as $\sim$130 ps, and triggered single-photon purity reaching $g^{(2)}(0) \sim 0.01$ under p-shell excitation. The fabrication yield reaches $\sim$50\% at sites where strain and e-beam-induced defects spatially overlap (Supplementary Note 9), confirming the scalability of the platform. The principal advance of this work is twofold: a reproducible and tunable MoTe$_2$ quantum emitter platform with state-of-the-art optical properties for TMD-based QEs, and the first demonstration of two-photon interference from this material system. Using a Hong–Ou–Mandel interferometer with a 12.5~ns delay line, we measured visibilities of $\sim$7.1\% (under p-shell excitation) and $\sim$3.6\% (under quasi-resonant excitation) within the full 12.5~ns integration window, rising to $>$30\% and $>$20\% at 200~ps post-selection, and reaching $\sim$60\% and $\sim$40\% after further temporal filtering. To the best of our knowledge, this represents the highest reported indistinguishability for any 2D TMD-based QE to date, and the first such demonstration for the MoTe$_2$ platform. 
We also investigated three complementary strategies for controlling the emitter properties: hBN encapsulation, in-plane electrical biasing, and optimized excitation through p-shell resonances. Encapsulation modifies the strain landscape from directional nanowrinkles to a more bubble-like morphology~\cite{androulidakis2018tailoring,Daveau2020Spectral}, leading to reduced confinement asymmetry, lower degree of linear polarisation (DOLP), and resolvable fine-structure splitting (FSS), in contrast to the larger FSS values (1.8–2.2~meV) observed for unencapsulated emitters under stronger wrinkle-induced strain. Electrical biasing enables continuous tuning over $\sim$3~meV -- sufficient to bring independent emitters into spectral resonance, a prerequisite for HOM interference between separate sources -- and additionally suppresses long-timescale blinking while reducing the radiative lifetime, lowering the linewidth-to-lifetime ratio to $R \sim 55$. This approaches the state-of-the-art for TMD-based QEs without cavity integration.
Optimized excitation through p-shell resonances identified by photoluminescence excitation (PLE) provides the third major control knob, suppressing background emission and enabling the high triggered single-photon purity reported above. \par
Compared to the only previous report on MoTe$_2$ quantum emitters~\cite{Zhao2021Site}, our emitters exhibit significantly narrower linewidths (150–300~\textmu eV vs. $\sim$600~\textmu eV) and much shorter lifetimes (130–450~ps vs. $>$20~ns). This difference likely arises from a stronger oscillator strength of our emitters, related to the directional strain profile and tighter confinement, rather than from non-radiative effects, as the PL intensity does not drop for the fastest emitters. The short lifetimes are consistent with the spin-allowed bright exciton ground state characteristic of Mo-based TMDs~\cite{Echeverry2016Splitting, Wang2018Colloquium}, similar to previous observations in MoSe$_2$ \cite{Yu2021Site}.
While the wavelength range of our emitters (1090–1200~nm) does not yet reach the telecom band, exploring strain engineering and dielectric environment control in thicker MoTe$_2$ flakes represents a promising direction toward telecom-band TMD single-photon sources, as reported by Zhao \textit{et al.}~\cite{Zhao2021Site}, who observed emitters up to 1550~nm in multilayer flakes. \par
The principal limitation of our current device is the modest two-photon interference visibility, set by fast decoherence on the picosecond–nanosecond timescale and reflected in coherence times of $\tau_c \sim$~60-80~ps. Importantly, the line broadening observed under second-scale spectroscopy ($\sim$150–300~\textmu eV) reflects slow charge wandering on timescales orders of magnitude longer than the HOM delay and, as shown in Section~VI, does not limit the indistinguishability. The strong dependence of visibility on temporal filtering is consistent with recent theoretical predictions for TMD-based emitters in the presence of phonon-induced and charge-noise-induced dephasing~\cite{Vannucci2024Single, Steinhoff2025Impact}.
Two complementary strategies are likely required to push indistinguishability beyond 50\%. First, integration of MoTe$_2$ QEs with high-Q photonic structures could exploit Purcell enhancement to shorten the radiative lifetime relative to the dephasing time and enhance the zero-phonon emission line~\cite{Iff2021Purcell, Sortino2021Bright, Flatten2018Microcavity, Luo2018Deterministic, Tripathi2018Spontaneous, Drawer2023Monolayer, Cai2018Radiative}. Second, advanced excitation schemes -- such as strict resonant excitation, swing-up of quantum emitter population (SUPER), polarization-selective excitation, or phonon-assisted excitation -- can further improve coherence and brightness~\cite{Kumar2016Resonant, Errando2021Resonance, vonHelversen2023Temperature, piccinini2024high, Vannucci2024Single, Wu2025Ultrahigh}. The reproducible fabrication process and active electrical control demonstrated here make MoTe$_2$ a strong candidate platform for both routes. \par
In summary, we report a reproducible approach for realizing high-quality, tunable quantum emitters in bilayer MoTe$_2$, exhibiting narrow linewidths, fast radiative decay, high single-photon purity, and the first measurable photon indistinguishability for the MoTe$_2$ platform. These results establish MoTe$_2$ as a viable candidate for scalable quantum photonic technologies in the near-infrared, and provide a foundation for further development toward telecom-band single-photon sources.

\bigskip
\section{Methods}

\subsection{Sample preparation}

\textit{Substrate fabrication:} The substrate consists of a distributed Bragg reflector (DBR) with alternating layers of GaAs and AlAs, fabricated by low-pressure metalorganic vapor-phase epitaxy (MOVPE). On top of the DBR, a $\sim$100~nm thick Al$_2$O$_3$ layer was deposited via atomic layer deposition (ALD) at $200~^\circ$C, providing electrical insulation and surface passivation. Alignment marks and electrodes were fabricated by spin coating and patterning UV resist (nLOF 2020, AZ), followed by physical vapor deposition of a Ti/Au/Ti multilayer (5/50/5~nm) and a liftoff process. \par
\textit{Nanopillar fabrication:} To fabricate the nanopillars, a small chip was cleaved from the wafer and spin-coated with high-resolution negative electron-beam resist (hydrogen silsesquioxane, HSQ; XR-1541-006, Dow Corning) at 2000~rpm for 1 minute. The coated sample underwent a two-step bake at $120~^\circ$C and $220~^\circ$C, each for 2 minutes. Nanopillars in the shape of $\sim$100~nm tall three-pointed stars (central diameter core: $\sim$0.4~\textmu{}m, tip-to-tip distance between arms: $\sim$4~\textmu{}m, length core-tip: $\sim$2.55~\textmu{}m), were patterned using e-beam lithography (JBX-9500FS, JEOL, 100~kV, 6~nA) with a dose of 11000~\textmu C/cm$^2$ and developed in a 1:3 AZ 400K:H$_2$O solution. \par
\textit{Exfoliation \& Transferring:} MoTe$_2$ flakes were exfoliated from bulk crystals (HQ Graphene) onto PDMS films prepared on glass slides, via the established scotch-tape technique \cite{Castellanos2014Deterministic}. Their thicknesses were confirmed via room-temperature photoluminescence (PL) under 650~nm LED excitation using an optical microscope (Nikon 50×, NA = 0.8). For the encapsulated samples, hexagonal Boron Nitride (hBN) flakes (5–15 layers) were prepared using the same exfoliation process and identified through an optical microscope (Nikon 20×, NA = 0.45). The exemplary hBN thickness measured by AFM in our devices is approximately 5.6–6.1~nm, corresponding to $\sim$10–15 layers, in the upper part of the indicated range. Flake transfer was performed using a transfer stage that could heat the substrate to $\sim$$70~^\circ$C, enabling van der Waals forces to overcome PDMS adhesion and release the flakes onto the substrate.\par
\textit{Defect fabrication:} Defects were introduced to the lattice by selective electron-beam bombardment (JBX-9500FS, JEOL, 100~kV, 1000~\textmu C/cm$^2$) of the prelocalized strained areas, enabling QEs formulation with increased fabrication yield (see Supplementary Note 9). \par
\textit{Sample reproducibility:} To investigate the reproducibility of the QEs' fabrication, additional samples were produced utilizing an identical sample fabrication method (more details in Supplementary Note 9). Moreover, the application of the full hBN encapsulation as well as the SiO$_2$/Si substrate was investigated to verify the potential impact of a different dielectric environment.

\subsection{Optical characterization}
All optical measurements of the samples were obtained with the help of a custom-built low-temperature micro-photoluminescence (\textmu PL) setup. The sample was mounted inside a closed-cycle cryostat (attoDRY800xs, Attocube) operating at a base temperature of $\sim$4~K. The cryostat was equipped with piezoelectric nanopositioners for precise spatial alignment and a high-numerical-aperture microscope objective (60×, NA = 0.8, LT-APO/Telecom, Attocube) integrated within the cryostat chamber to facilitate efficient collection of the photoluminescence signal. Excitation of the quantum emitters was performed using two types of semiconductor diode lasers (PicoQuant): an above-band laser at 650~nm (LDH-D-C-650) and a quasi-resonant laser at 1080~nm (LDH-D-C-1080). Both lasers operated in continuous-wave (CW) or pulsed modes, delivering pulses shorter than 100~ps with adjustable repetition rates ranging from 2.5 to 80~MHz. Photoluminescence excitation spectroscopy and p-shell excitation were performed by utilizing a tunable femtosecond laser (Coherent Chameleon Ultra II) combined with an optical parametric oscillator (Coherent Chameleon Compact OPO-Vis) and laser spectral filtering from a home-built pulse shaper equipped with a 600~lines/mm diffraction grating. \textmu PL spectra were acquired using a spectrometer based on a 0.328~m focal-length monochromator (Kymera 328i, Andor, Oxford Instruments) equipped with interchangeable gratings (150, 600, and 830~lines/mm) and a deep thermo-electrically cooled (In,Ga)As linear array detector (iDus 1.7~\textmu m, Andor, Oxford Instruments), providing a maximum spectral resolution of $\sim$140-160 \textmu eV. Time-resolved PL characterization and single-photon statistics measurements were performed via time-correlated single-photon-counting mode. To verify single-photon emission, a Hanbury Brown and Twiss (HBT) configuration employing a 50:50 fiber beamsplitter was used. Photon indistinguishability was assessed via Hong–Ou–Mandel (HOM) interference, utilizing a free-space beamsplitter (Wollaston prisms), a half-wave plate for polarization rotation in one arm, a second fiber-based beamsplitter, and a tunable optical delay line implemented via a combination of fiber and a linear translation stage. A three-paddle fiber polarization controller was employed for interferometer calibration. Detection was carried out using superconducting nanowire single-photon detectors (SNSPDs; ID281, ID Quantique) connected to a high-resolution single-photon counting module (Time Tagger Ultra, Swabian Instruments). Polarization-resolved PL measurements were performed using a rotating half-wave plate followed by a linear polarizer.

\subsection{Optical simulation}
To estimate the extraction efficiency presented in Fig.~\ref{fig:one}b, we model the QE as a classical point dipole with in-plane orientation and harmonic time dependence at frequency $\omega$. The corresponding current density is $\mathbf{J}(\mathbf{r})=-i\omega\mathbf{p}\delta(\mathbf{r}-\mathbf{r}_{\rm d})$, where $\mathbf{r}_{\rm d}$ is the position of the QE and $\mathbf{p}$ is the dipole moment. We then use the method described in the Supplementary Information of Ref.\ \cite{Wyborski2025a} to solve Maxwell's equations in the frequency domain.

\bigskip

\section*{Acknowledgements}

The authors acknowledge support from the European Research Council (ERC-StG "TuneTMD", grant no. 101076437) and the Villum Foundation (grant no. VIL53033). CR acknowledges the Danish National Research Foundation (grant no. DNRF147 NanoPhoton). BM and CR acknowledge the European Union's Horizon Europe Research and Innovation Programme under the project QPIC 1550 (grant no. 101135785). The authors acknowledge E. Semenova for her support in the growth of the substrate. The authors also acknowledge the European Research Council (ERC-CoG "Unity", grant no. 865230) and the cleanroom facilities at the Danish National Centre for Nano Fabrication and Characterization (DTU Nanolab).

\section*{Author contributions}

AP and PM fabricated the MoTe$_2$ quantum emitters on the substrates and processed the samples.
CR grew the substrate, including a distributed Bragg reflector.
PW, AP and PM performed the optical characterization of quantum emitters. 
PW performed the data analysis and processing of all the data.
MJ and NG performed calculations on the optical properties of the photonic structure.
BM supervised and coordinated the project. 
PW, AP, PM, and BM wrote the manuscript with the input of all co-authors.

\section*{Competing Interests}
The authors declare no competing interest.

\appendix

\bibliography{biblio.bib}
\end{document}